\pdfoutput=1

\documentclass[11pt]{article}

\usepackage[T1]{fontenc}
\usepackage{lmodern}
\usepackage{microtype}
\usepackage{geometry}
\usepackage{amsmath,amssymb,amsthm}

\theoremstyle{definition}

\usepackage{graphicx}
\usepackage{booktabs}
\usepackage{tabularx}
\usepackage{longtable}
\usepackage{multirow}
\usepackage{siunitx}
\usepackage{subcaption}

\usepackage{algorithm}
\usepackage{algpseudocode}

\usepackage{tikz}
\usepackage{forest}
\usetikzlibrary{positioning,calc,arrows.meta}
\newlength\flowW
\tikzset{
  flow/node width/.store in=\flowNodeW,
  flow/node height/.store in=\flowNodeH,
  flow/hsep/.store in=\flowHsep,
  flow/vsep/.store in=\flowVsep,
}
\tikzset{
  flow/node width=.32\flowW,
  flow/node height=9mm,
  flow/hsep=.06\flowW,
  flow/vsep=3mm,
}

\usepackage{hhline}
\usepackage{array}
\usepackage{makecell}
\usepackage{pifont}      
\usepackage{amssymb}     
\usepackage{tabularx}

\newcolumntype{L}[1]{>{\raggedright\arraybackslash}p{#1}}
\newcolumntype{C}[1]{>{\centering\arraybackslash}p{#1}}
\newcolumntype{R}[1]{>{\raggedleft\arraybackslash}p{#1}}
\newcolumntype{Y}{>{\centering\arraybackslash}X} 

\usepackage{hyperref}
\usepackage[nameinlink,capitalize,noabbrev]{cleveref}
\hypersetup{
  colorlinks=true,
  linkcolor=blue,
  citecolor=blue,
  urlcolor=blue,
  pdftitle={TITLE},
  pdfauthor={AUTHOR LIST}
}

\usepackage[backend=bibtex,style=nature]{biblatex}
\title{The Vienna 4G/5G Drive-Test Dataset}

\author{
  Wilfried Wiedner$^{1,2,*}$ \and
  Lukas Eller$^{1,2,\dag}$ \and
  Mariam Mussbah$^{1,2}$ \and
  Dominik R\"ossler$^{1,2}$ \and
  Valerian Maresch$^{1,2}$ \and
  Philipp Svoboda$^{1,2}$ \and
  Markus Rupp$^{1}$
}

\date{} 

\begin{document}
\maketitle

\begin{center}
{\small
$^{1}$Institute of Telecommunications, TU Wien, Gusshausstraße 25, 1040 Vienna, Austria\\
$^{2}$Christian Doppler Laboratory for Digital Twin assisted AI for sustainable Radio Access Networks, TU Wien\\
$^{*}$Corresponding Author(s): Wilfried Wiedner (\texttt{wilfried.wiedner@tuwien.ac.at})
}
\end{center}

\begin{abstract}
\noindent Machine learning for mobile network analysis, planning, and optimization is often limited by the lack of large, comprehensive real-world datasets. This paper introduces the Vienna~4G/5G Drive-Test Dataset, a city-scale open dataset of georeferenced Long Term Evolution (LTE) and 5G New Radio (NR) measurements collected across Vienna, Austria. The dataset combines passive wideband scanner observations with active handset logs, providing complementary network-side and user-side views of deployed radio access networks. The measurements cover diverse urban and suburban settings and are aligned with time and location information to support consistent evaluation. For a representative subset of base stations (BSs), we provide inferred deployment descriptors, including estimated BS locations, sector azimuths, and antenna heights. The release further includes high-resolution building and terrain models, enabling geometry-conditioned learning and calibration of deterministic approaches such as ray tracing. To facilitate practical reuse, the data are organized into scanner, handset, estimated cell information, and city-model components, and the accompanying documentation describes the available fields and intended joins between them. The dataset enables reproducible benchmarking across environment-aware learning, propagation modeling, coverage analysis, and ray-tracing calibration workflows.
\end{abstract}


\renewcommand{\thefootnote}{\dag}
\footnotetext{This work was conducted while Lukas Eller was with the Institute of Telecommunications, TU Wien, Austria. He is now with Huawei Technologies, Munich, Germany.}

\section*{Background \& Related Work}
Machine learning is increasingly applied to mobile network analysis, planning, and optimization, yet progress is constrained by the limited availability of comprehensive real-world datasets. The gap is most severe for geometry- and environment-conditioned problems, including radio map estimation and coverage prediction~\cite{radio_map_estimation}, propagation modeling and calibration~\cite{dl_nw_planer_eller}, ray-tracing simulation (\textit{e.g.}, NVIDIA Sionna~RT)~\cite{sionna_rt}, measurement-based base-station localization \cite{eller_localization}, fingerprinting and channel charting~\cite{channel_charting}, digital network twin (DNT) construction and calibration~\cite{digital_twin}, and 5G beam-level analysis and beam management~\cite{beam_paper}.

Beyond enabling learning-based methods, modern network engineering workflows also require datasets that support reproducible benchmarking against established model-based approaches. In particular, propagation analysis and deterministic modeling (\textit{e.g.}, ray tracing) rely on calibration measurements together with explicit deployment descriptors and detailed 3D environmental context. Datasets that pair real-world measurements with both cell metadata and building/terrain information therefore provide a strong foundation for benchmarks spanning artificial intelligence (AI)-based and traditional methods, including environment-aware learning and digital twin (DT) calibration.

Several open measurement datasets have recently been released, providing valuable empirical evidence on 4G/5G coverage and performance across different cities and scenarios~\cite{rochman_dataset,kousias_dataset,farthofer_open_dataset}. However, these datasets typically focus on measurement traces and key performance indicators (KPIs), and do not jointly provide explicit deployment descriptors together with detailed building/terrain context.

Complementing these real-world measurement efforts, \citeauthor{yapar2022dataset} \cite{yapar2022dataset} propose a synthetic dataset of pathloss and time-of-arrival maps generated via ray tracing on OpenStreetMap-based city models. While it incorporates detailed building and terrain geometry and is well suited for propagation-modeling studies, its ray-traced nature may not fully capture the complexity of real operator deployments or measured network performance.

A more recent ray-tracing-based benchmark for radio-map learning is introduced in \cite{jaensch_radiomap_dataset}, offering a large collection of simulated radio maps alongside map-derived features such as directional transmitter antennas and environment descriptors derived from building-height and vegetation models. While valuable for benchmarking radio-map estimation methods, it remains entirely synthetic and likewise lacks real operator deployments and measurement-based validation; indeed, comparisons between ray-tracing simulations and field measurements can show substantial discrepancies~\cite{fastenbauer}.

More recent large-scale empirical datasets have further expanded the landscape of open mobile network measurements, including multi-campaign collections for Long Term Evolution (LTE) and 5G New Radio (NR) analysis and rich drive-test datasets aimed at quality-of-service (QoS) prediction and mobility-aware modeling~\cite{chronicals_5gnsa,nordicdat_2024}. While these datasets significantly enrich the ecosystem of open mobile network data, they still focus primarily on performance metrics and mobility traces rather than on the explicit integration of deployment descriptors and high-resolution environmental information.

Among the existing open datasets, the \emph{DoNext} dataset~\cite{dortmund2025dataset} represents the closest empirical point of comparison to the dataset presented in this work. DoNext comprises a large-scale collection of 4G and 5G measurements from Dortmund, Germany, covering mobile, static, and rail-based campaigns over a period of two years. While it provides extensive active and passive measurements and estimated base station (BS) assignments, it does not include explicit cell deployment information or high-resolution environmental data such as terrain and building models.

Taken together, these efforts demonstrate substantial progress toward open mobile network benchmarking. However, large-scale datasets that jointly combine \emph{real-world} measurements with deployment descriptors and high-resolution building/terrain context remain scarce---despite being central to environment-aware modeling and DT construction. To close this gap, we present the Vienna~4G/5G Drive-Test Dataset, a large-scale open-access resource that combines city-wide drive-test measurements with inferred deployment descriptors and a building/terrain model of Vienna. \autoref{tab:datasets} situates the Vienna~4G/5G Drive-Test Dataset~\cite{wiedner_vienna_2025} within the landscape of publicly available datasets, comparing modality, scale, and available context. Below, we summarize the dataset contents and representative use cases it enables, including systematic evaluation and benchmarking.

\begin{table}[!t]
\caption{Comparison of publicly available datasets. Digital twin readiness refers to the joint availability of deployment metadata (\textit{e.g.}, BS positions, orientations) and environmental context (\textit{e.g.}, buildings, terrain). Multi-MNO indicates measurements from multiple mobile network operators (MNOs); A+P denotes active and passive collection.}
\label{tab:datasets}
\renewcommand{\arraystretch}{1.2}
\centering
\setlength{\tabcolsep}{4pt}

\begin{tabularx}{\linewidth}{|>{\raggedright\arraybackslash}X|c|c|c|c|c|}
\hline
\textbf{Dataset} & \textbf{Type} & \textbf{Multi-} & \textbf{Active/} & \textbf{Scale} & \textbf{Digital} \\
                & 4G/5G         & \textbf{MNO}    & \textbf{Passive} &                & \textbf{Twin} \\
\hline
\citeauthor{rochman_dataset} \cite{rochman_dataset}          & \checkmark/\checkmark & \checkmark & Active & 2 Cities, Highway & \ding{55} \\
\citeauthor{kousias_dataset} \cite{kousias_dataset}          & \checkmark/\checkmark & \checkmark & A+P    & City              & \ding{55} \\
\citeauthor{farthofer_open_dataset} \cite{farthofer_open_dataset} & \checkmark/\checkmark & \ding{55}  & Active & Highway           & \ding{55} \\
\citeauthor{yapar2022dataset} \cite{yapar2022dataset}           & Synthetic             & --         & --     & City              & Env.~only \\
\citeauthor{jaensch_radiomap_dataset} \cite{jaensch_radiomap_dataset} & Synthetic             & --         & --     & City              & Env.~only \\
\citeauthor{nordicdat_2024} \cite{nordicdat_2024}          & \checkmark/\checkmark & \ding{55}  & Active & Cities, Borders   & \ding{55} \\
\emph{DoNext}~\cite{dortmund2025dataset}       & \checkmark/\checkmark & \checkmark & A+P    & City              & \ding{55} \\
\hline
\textbf{Vienna 4G/5G (This work)} & \checkmark/\checkmark & \checkmark & A+P & City & \textbf{\checkmark~(net+env)} \\
\hline
\end{tabularx}
\end{table}

\subsection*{The Vienna~4G/5G Drive-Test Dataset at a glance.}
\begin{itemize}
  \item \textbf{Scale and diversity:} city-wide drive-test measurements with dense geospatial sampling across suburban and urban areas.
  \item \textbf{Dual modality:} passive scanner + active handset logs (network-side + user-side).
  \item \textbf{3D environmental context:} building and terrain models enabling environment-aware workflows.
  \item \textbf{Deployment descriptors:} for a subset of sites, we provide estimated site locations, sector azimuths, and antenna heights, enabling deployment-aware analyses.
  \item \textbf{LTE and 5G~NR (NSA):} consistent KPI set across radio access technologies (RATs), including NR beam-level measurements for benchmarking and beam-management studies.
\end{itemize}

\begin{figure}[t]
  \centering
  \includegraphics[height=6cm,keepaspectratio]{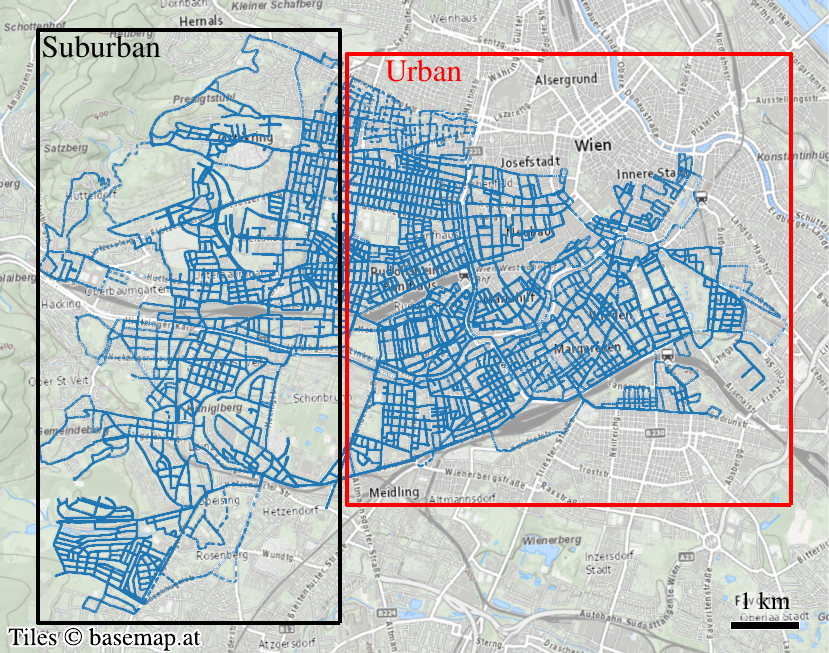}
  \caption{City-wide measurement coverage with dense sampling across suburban and urban areas (indicated by the black and red boxes).}
  \label{fig:overview}
\end{figure}

\begin{figure*}[t]
  \centering
  \includegraphics[width=\textwidth]{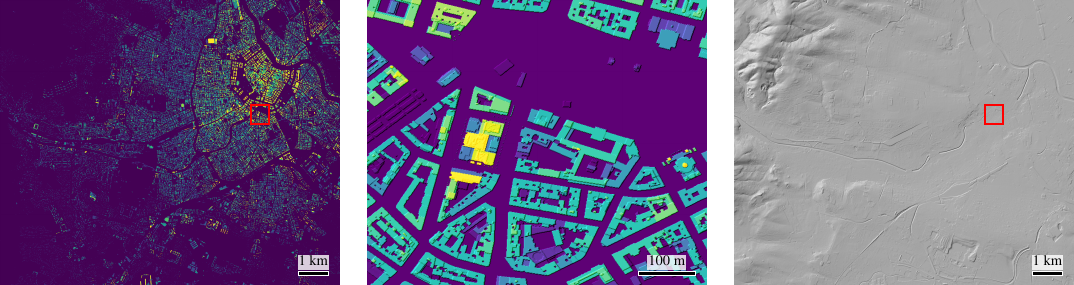}
  \caption{Overview of the city model provided with the dataset: building model (left), a zoomed-in building region of interest (center, indicated by the red box), and terrain model overview (right).}
  \label{fig:city_model}
\end{figure*}

\begin{figure*}[t]
  \centering
  \includegraphics[width=\textwidth]{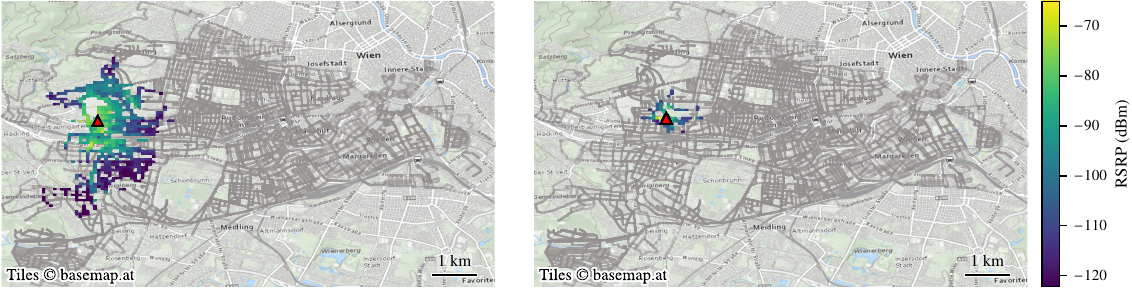}
  \caption{Example RSRP coverage maps around two BSs with different antenna heights: 38\,m (left) and 19\,m (right). The red triangle marks the estimated BS position; colors show RSRP.}
  \label{fig:bs_heights}
\end{figure*}

\begin{figure*}[t]
  \centering
  \includegraphics[width=\textwidth]{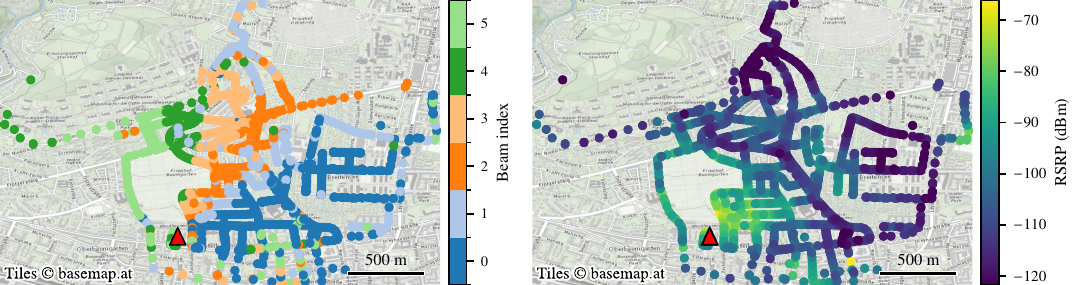}
    \caption{Example NR beam-level view for a single cell of the corresponding BS: strongest beam index per location (left) and the corresponding RSRP (right). The BS position is marked by a red triangle.}
  \label{fig:beam_rsrp}
\end{figure*}

\noindent Figures \ref{fig:overview}--\ref{fig:beam_rsrp} provide representative examples of the dataset components summarized above. To improve readability, \autoref{tab:abbreviations} summarizes the main abbreviations used throughout the manuscript.

\subsection*{Representative evaluation tasks and benchmarking use cases enabled by the Vienna~4G/5G Drive-Test Dataset.}
\begin{itemize}
  \item \textbf{Ray-tracing calibration and benchmarking:} use measured links together with deployment descriptors and 3D building/terrain context to calibrate deterministic simulators and benchmark ray-tracing predictions against real-world observations.
  \item \textbf{Environment-aware radio map learning:} train and evaluate geometry-conditioned models using building/terrain features, and compare against classical and deterministic baselines.
  \item \textbf{Deployment inference and validation:} study site localization and sector-parameter estimation (\textit{e.g.}, azimuth/height) from measurement fingerprints and inferred deployment descriptors.
  \item \textbf{Beam-management studies:} evaluate NR beam selection and tracking under mobility using beam-level measurements.
  \item \textbf{Digital twin prototyping and calibration:} combine deployment metadata, 3D city model, and measurements for city-scale DT workflows.
\end{itemize}

\begin{longtable}{|l|l|}
\caption{List of abbreviations used in this manuscript.\label{tab:abbreviations}}\\
\hline
\textbf{Abbreviation} & \textbf{Meaning} \\
\hline
\endfirsthead

\hline
\textbf{Abbreviation} & \textbf{Meaning} \\
\hline
\endhead
AI & Artificial Intelligence \\
BKM & Building Model \\
BS & Base Station \\
CINR & Carrier-to-Interference-plus-Noise Ratio \\
CRS & Cell-specific Reference Signal \\
DNT & Digital Network Twin \\
DOI & Digital Object Identifier \\
DT & Digital Twin \\
ECDF & Empirical Cumulative Distribution Function \\
eNB & eNodeB \\
EPSG & European Petroleum Survey Group geodetic parameter registry \\
GLM & Ground-Level Model \\
gNB & gNodeB \\
GPS & Global Positioning System \\
KPI & Key Performance Indicator \\
LTE & Long Term Evolution \\
MIB & Master Information Block \\
MNO & Mobile Network Operator \\
NR & New Radio \\
NSA & Non-Standalone \\
PCell & Primary Cell \\
PCI & Physical Cell Identity \\
PDSCH & Physical Downlink Shared Channel \\
PUSCH & Physical Uplink Shared Channel \\
QoS & Quality of Service \\
RAT & Radio Access Technology \\
RMSE & Root Mean Square Error \\
RSRP & Reference Signal Received Power \\
RSRQ & Reference Signal Received Quality \\
RSSI & Received Signal Strength Indicator \\
SINR & Signal-to-Interference-plus-Noise Ratio \\
SIB & System Information Block \\
PSCell & Serving Primary Secondary Cell \\
SSB & Synchronization Signal Block \\
TA & Timing Advance \\
ToA & Time of Arrival \\
UE & User Equipment \\
UTC & Coordinated Universal Time \\
\hline
\end{longtable}

\section*{Methods}

The dataset was collected between June~2024 and March~2025 and covers approximately $\SI{100}{\square\kilo\metre}$ of Vienna, spanning both urban and suburban environments. To clarify the temporal extent of the measurement campaign, \autoref{tab:temporal_summary} summarizes the distribution of released measurement records by measurement day. Records denote rows in the released
measurement tables, and measurement IDs group records with the same operator and timestamp within each table. Measurements were collected during discrete drive-test runs rather than according to a fixed weekly or monthly repetition schedule. Routes were selected to obtain broad and dense spatial coverage across urban and suburban parts of Vienna, not to intentionally repeat measurements at
the same locations. Nevertheless, some spatial overlap between runs occurs because routes may intersect or traverse the same road segments on different measurement days.

\begin{table}[t!]
\centering
\small
\caption{Temporal distribution of the released measurement records by measurement
day. Locations are counted as unique latitude/longitude pairs after rounding to
five decimal places. Cells/PCIs denote unique valid cell identifiers where available; otherwise, unique PCIs are counted instead.}
\label{tab:temporal_summary}
\begin{tabular}{|l|r|r|r|r|}
\hline
\textbf{Date} & \textbf{Records} & \textbf{Meas. IDs} & \textbf{Locations} & \textbf{Cells/PCIs} \\
\hline
2024-06-27 & $245\,422$ & $29\,665$ & $6\,397$ & $892$ \\
2024-07-01 & $180\,808$ & $20\,296$ & $7\,420$ & $726$ \\
2024-07-04 & $76\,750$ & $7\,697$ & $3\,105$ & $617$ \\
2024-07-05 & $97\,409$ & $8\,446$ & $3\,517$ & $717$ \\
2024-08-22 & $19\,318$ & $4\,025$ & $2\,060$ & $42$ \\
2024-08-23 & $160\,774$ & $26\,106$ & $14\,029$ & $232$ \\
2024-08-25 & $232\,602$ & $37\,442$ & $18\,519$ & $578$ \\
2024-08-26 & $178\,093$ & $29\,424$ & $14\,194$ & $374$ \\
2024-08-27 & $34\,287$ & $5\,492$ & $2\,516$ & $189$ \\
2024-08-28 & $186\,089$ & $28\,145$ & $14\,193$ & $399$ \\
2024-08-30 & $193\,371$ & $30\,825$ & $15\,427$ & $642$ \\
2024-10-13 & $86\,644$ & $40\,826$ & $18\,893$ & $1\,047$ \\
2024-10-14 & $95\,744$ & $45\,211$ & $20\,572$ & $688$ \\
2024-10-26 & $144\,398$ & $102\,788$ & $20\,551$ & $1\,068$ \\
2024-10-27 & $419\,181$ & $245\,212$ & $62\,497$ & $3\,027$ \\
2024-10-28 & $158\,850$ & $152\,794$ & $32\,801$ & $765$ \\
2025-02-08 & $299\,485$ & $138\,227$ & $14\,392$ & $1\,940$ \\
2025-03-08 & $83\,904$ & $34\,077$ & $4\,051$ & $406$ \\
\hline
Total & $2\,893\,129$ & $986\,698$ & $929\,217$ & $7\,911$ \\
\hline
\end{tabular}
\end{table}

\begin{figure}
\centering
\includegraphics[width=\linewidth]{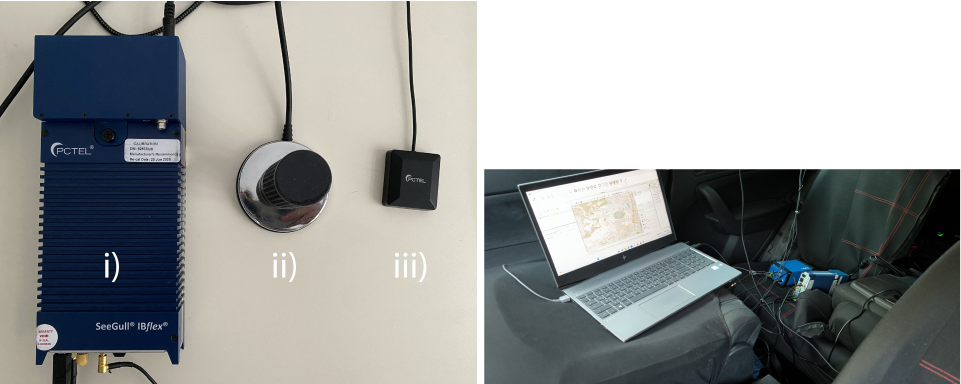}
\caption{Equipment for passive measurements (left) consisting of the scanner (i), two OmniLOG~PRO antennas (ii), and a GPS receiver (iii). Scanner setup in the car (right).}
\label{fig:equipment}
\end{figure}

Passive measurements were performed using a PCTEL IBflex scanning receiver (i), two OmniLOG PRO omnidirectional antennas (ii) mounted on the car rooftop, and a PCTEL Global Positioning System (GPS) receiver (iii), as illustrated in \autoref{fig:equipment}. The scanner was connected to a notebook running Nemo Outdoor software from Keysight (\autoref{fig:equipment}), which was used to control the measurement process and record the data. This setup enabled wideband, technology-agnostic monitoring of cellular signals with spatial diversity, providing an unbiased view of the deployed networks across the measurement area.

In parallel, active phone measurements were performed using three 5G-capable smartphones mounted on the vehicle dashboard and connected to Nemo Outdoor, capturing radio-quality metrics such as reference signal received power (RSRP), reference signal received quality (RSRQ), received signal strength indicator (RSSI), and signal-to-interference-plus-noise ratio (SINR), as well as user-centric performance indicators including throughput and timing advance (TA).

At each georeferenced measurement point (measurement ID/timestamp + GPS position), the dataset records one or more cell observations together with their identifiers—most notably PCI and channel number (and, where available, cell/BS/sector IDs)—and radio-layer KPIs. Scanner logs provide passive multi-cell (and for NR, beam-level) observations, whereas phone logs primarily report serving-cell measurements and add user-centric fields such as throughput and TA; in non-standalone (NSA) 5G NR phone logs, physical NR cell/gNB identifiers are not exposed (optional dataset-internal dummy IDs may be provided).

Although the underlying devices support considerably higher measurement rates, the effective temporal granularity of the recorded reference-signal measurements is approximately 500\,ms for both scanner and handset data. This granularity is determined by the Nemo Outdoor recording process and therefore does not directly correspond to the nominal measurement rates of the individual devices.

\subsection*{Scanner Data Processing}

Post-processing followed two parallel pipelines for scanner and phone data. For the scanner measurements, raw Nemo outputs were parsed to identify master information block (MIB) and system information block (SIB) decoding events, as well as cell-specific reference signal (CRS) and synchronization signal block (SSB) events. Multi-valued records containing several physical cell identifiers (PCIs), channels, or beams were exploded into individual entries, standardized, and cleaned by filtering incomplete data.

CRS and SSB measurements were subsequently merged with the corresponding MIB and SIB information by first identifying candidate pairs via exact keys (PCI, channel, file) and then resolving ambiguities through temporal alignment and spatial filtering. The resulting records were enriched with deployment metadata such as eNodeB (eNB) ID and sector ID (LTE).

Raw Nemo logs contain hundreds of parameters in nested and multi-valued formats, with partially overlapping but RAT-specific structures. We therefore provide harmonized, analysis-ready scanner tables that expose a consistent set of scanner-side measurements (CRS/SSB events with associated MIB/SIB context) and preserve temporal and spatial alignment at scale.

\subsection*{Phone Data Processing}

For the phone data, post-processing focused on LTE primary cells (PCell) and NR serving primary secondary cells (PSCell), which form the basis of all handset-based measurements in the released dataset. A harmonized subset of key measurements---RSRP, RSRQ, RSSI, SINR, physical downlink shared channel (PDSCH) throughput, physical uplink shared channel (PUSCH) throughput, and TA---was extracted to ensure consistency across records. In the current release, handset-based 5G~NR measurements in the phone data table are available for one operator only.

The reference-signal and radio-quality measurements in the released handset data have an effective temporal granularity of approximately 500\,ms, as determined by the Nemo Outdoor recording process. Other handset KPIs are reported less regularly and may therefore be unavailable at individual timestamps. In particular, PDSCH/PUSCH throughput values are only available when corresponding shared-channel activity is reported by the diagnostic interface, while TA is reported at an even lower frequency.

During the drive-test campaign, the RTR-Netztest~\cite{rtr_netztest} application was operated in loop mode to generate repeated ordinary speed-test traffic while the vehicle was moving. The loop mode was configured with the maximum configurable number of 500 tests, while the waiting time and distance between successive tests were set to their minimum configurable values of five minutes and 25\,m, respectively. These settings maximized user-plane traffic during the measurement runs and provided a practical and reproducible source of data activity without requiring separate dedicated downlink- and uplink-only measurement campaigns. Radio-layer information was logged in parallel from the handsets using the Qualcomm diagnostic interface. Consequently, the throughput fields released in the phone tables represent physical-layer shared-channel throughput indicators rather than application-layer throughput values reported by the speed-test application. Specifically, downlink throughput corresponds to PDSCH throughput and uplink throughput to PUSCH throughput as reported in the diagnostic logs.

To prepare TA for downstream use, duplicate timestamps were merged, missing coordinates interpolated, and invalid values discarded. The resulting TA records were then matched with the corresponding PCell or PSCell measurements based on PCI and channel number, selecting the closest spatial candidates to ensure temporal and spatial consistency. These TA-based records were subsequently used for the base-station location estimates reported in this work.

\subsection*{Base Station Estimation and Digital Network Twin Construction}

To support the creation of a DNT, the locations of a representative subset of BSs were estimated for each radio access technology (RAT) and operator using the methodology of \citeauthor{eller_localization}~\cite{eller_localization}. Public BS entries from the Austrian Senderkataster~\cite{senderkataster} were used to constrain the search space, ensuring that only physically plausible candidate sites were considered. Antenna heights were subsequently estimated using Google Earth Pro.

For LTE, the selection of BSs was performed on a per-day basis by identifying the ten most frequently observed BSs per operator for each measurement day. This resulted in 50 distinct LTE BSs per operator across the full measurement campaign. This strategy was chosen to ensure broader spatial and temporal coverage of the network, rather than concentrating the estimation on a small number of persistently dominant sites. As LTE measurements include network-reported cell identifiers, measurements could be directly grouped by BS prior to location estimation.

For 5G NR measurements, the handset logs collected in NSA mode do not expose physical NR cell or gNodeB (gNB) identifiers, which prevents direct grouping of measurements belonging to a specific BS. To address this limitation, we adopt a PCI-based grouping heuristic. Specifically, we assume that within a single measurement day a given NR PCI is not reused across different gNBs, which is reasonable in our setting given the limited geographic coverage per day and the typically large NR PCI space. Based on this assumption, the thirty most frequently observed PCIs per measurement day were selected, subject to the additional requirement that each PCI be represented by at least five measurement samples acquired at distinct locations.

Prior to BS localization, spatial consistency was assessed for each selected PCI by constructing TA circles centered at the corresponding user equipment (UE) measurement locations, with the TA value defining the circle radius. The overlap of these circles was evaluated to determine whether the measurements could plausibly originate from a common transmitter location. In cases where the TA circles associated with the same PCI formed disjoint spatial subsets---indicating multiple spatially inconsistent transmitter hypotheses---only the subset containing the largest number of samples was retained for subsequent BS location estimation. This procedure ensures that the localization step is applied only to the most spatially consistent and well-supported measurement cluster. Due to data availability constraints, this 5G NR cell estimation procedure is currently applied for one network operator only.

Finally, sector orientations were estimated using handset-based measurements. For each estimated BS, the RSRP-weighted centroid of the associated phone measurement locations was computed, and the azimuth from the estimated BS position to this centroid was calculated. An azimuth of \SI{0}{\degree} corresponds to geographic north, with angles increasing clockwise, consistent with standard cartographic and navigation conventions~\cite{esri_gis,winncom_azimuth}. Phone measurements were used for sector orientation estimation to ensure methodological consistency with the timing-advance--based localization process and to avoid ambiguities arising from beam- or snapshot-level scanner observations.

\begin{figure}
\centering
\includegraphics[width=0.6\textwidth]{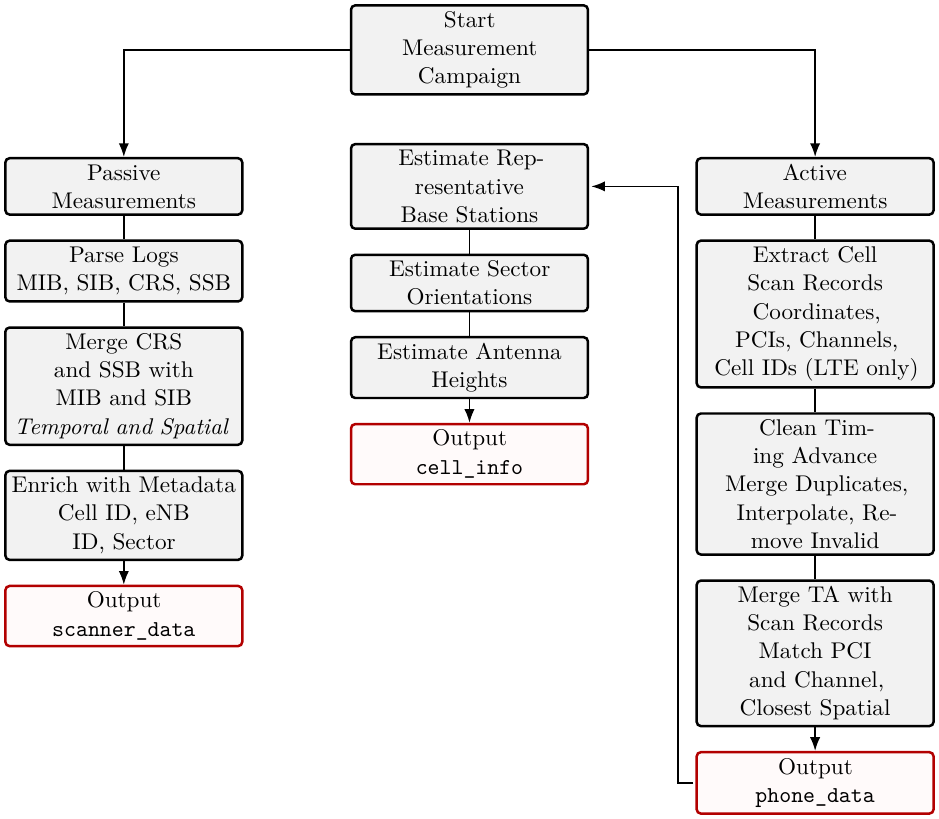}
\caption{Illustration of the data collection and processing pipeline.}
\label{fig:collection_processing}
\end{figure}

The overall data collection and processing chain is illustrated in \autoref{fig:collection_processing}, where TA is depicted as the representative phone-side performance indicator; the released dataset additionally includes RSRP, RSRQ, RSSI, SINR, and throughput.

\section*{Data Record}

\forestset{
  folder/.style={
    draw,
    rounded corners,
    fill=yellow!70,
    font=\ttfamily,
    minimum height=4ex,
    inner sep=2pt,
  },
  file/.style={
    draw,
    fill=white,
    font=\ttfamily,
    minimum height=3.5ex,
    inner sep=2pt,
  }
}

\begin{figure}[!t]
\centering
\includegraphics[]{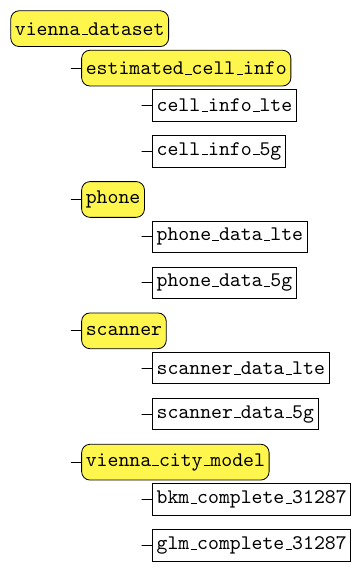}
\caption{Directory structure of the Vienna~4G/5G Dataset.}
\label{fig:dataset_structure}
\end{figure}

\noindent The Vienna 4G/5G Drive-Test Dataset is deposited in Zenodo and is available 
under the versioned DOI 
\url{https://doi.org/10.5281/zenodo.21372657}~\cite{wiedner_vienna_2025}. 
The dataset is also archived under the concept DOI 
\url{https://doi.org/10.5281/zenodo.18322065}, 
which resolves to all released versions, including future updates. 
This article describes version 5 of the dataset.

As illustrated in \autoref{fig:dataset_structure}, the Vienna~4G/5G Drive-Test
Dataset is organized into four main subdirectories: (i) estimated cell
information, (ii) phone measurements, (iii) scanner measurements, and (iv) the
Vienna city model. \autoref{tab:dataset_summary} provides a concise overview
of the size and coverage of the released measurement tables. Measurement and
metadata files are provided in CSV and Parquet formats, while terrain and
building models are distributed as GeoTIFF rasters. The combination of estimated
cell metadata (BS locations, antenna heights, and sector azimuths) with the
high-resolution city model enables environment-aware analyses and learning
workflows, such as training propagation surrogates and supporting AI-assisted
network planning. The following subsections describe the contents and available
columns in each subdirectory. Complementary technical notes and ancillary
information are available in the associated repository.

\begin{table}[!t]
\centering
\caption{Summary statistics of the released measurement tables. Records denote
rows in the released CSV/Parquet files. Measurement IDs are file-local
operator-time groups. Locations are counted as unique latitude/longitude pairs
after rounding to five decimal places. Cells/PCIs denote unique valid cell identifiers where available; for handset NR, unique PCIs are reported because physical NR cell identifiers are not available in NSA mode.}
\label{tab:dataset_summary}
\begin{tabular}{|l|r|r|r|r|}
\hline
\textbf{File} & \textbf{Records} & \textbf{Meas. IDs} & \textbf{Locations} & \textbf{Cells/PCIs} \\
\hline
\texttt{scanner\_data\_lte} & $529\,724$ & $193\,457$ & $129\,859$ & $5\,679$  \\
\texttt{scanner\_data\_5g}  & $1\,602\,718$  & $141\,873$  & $79\,474$  & $1\,371$  \\
\texttt{phone\_data\_lte}   & $596\,543$ & $595\,470$ & $136\,864$ & $2\,948$ \\
\texttt{phone\_data\_5g}    & $164\,144$ & $55\,898$ & $26\,405$ & $273$ \\
\hline
Total & $2\,893\,129$ & $986\,698$ & $255\,090$ & $7\,911$ \\
\hline
\end{tabular}
\end{table}

\subsection*{Common Metadata and Identifiers}

Several identifiers are used to enable spatial, temporal, and operator-level
alignment between measurement records and estimated cell metadata. These fields
are shared across scanner and phone measurements and are summarized in
\autoref{tab:common_metadata}. They enable merging per-sample measurements
with per-cell metadata, which is essential for environment-aware propagation
learning and planning-oriented machine learning models.

For LTE, standardized network-reported identifiers
(\texttt{cell\_id}, \texttt{enb\_id}, \texttt{sector\_id}) are available and used
directly throughout the dataset. For 5G NR measurements collected in
NSA mode, physical NR cell and gNB (5G BS) identifiers are not exposed by
commercial handsets. In such cases, auxiliary dataset-internal identifiers
(\texttt{cell\_id\_dummy} and \texttt{gnb\_id\_dummy}) are provided where reliable
cell estimates are available.

These auxiliary identifiers do not correspond to standardized 3GPP identifiers.
They serve exclusively as stable internal keys to enable grouping of NR
measurements and linkage to estimated NR cell metadata. Their presence is
optional, and the corresponding fields may be null for NR measurements where no
cell estimates exist.

\begin{table}[!t]
\caption{Common metadata fields across all measurement types.}
\label{tab:common_metadata}
\renewcommand{\arraystretch}{1.2}
\centering
\setlength{\tabcolsep}{3pt}
\begin{tabular}{|L{90pt}|L{40pt}|L{35pt}|L{275pt}|}
\hline
\textbf{Column} & \textbf{Type} & \textbf{Unit} & \textbf{Description} \\
\hline
operator & String & -- & Mobile network operator identifier (A, B, or C) \\
channel\_number & Integer & -- & Radio frequency channel number \\
frequency\_khz & Integer & kHz & Carrier center frequency \\
pci & Integer & -- & Physical Cell Identity \\
cell\_id & Integer & -- & Unique cell identifier (LTE and NR where available) \\
enb\_id / gnb\_id & Integer & -- & BS identifier (LTE/NR where available) \\
cell\_id\_dummy & Integer & -- & Auxiliary dataset-internal cell identifier (NR only, where available) \\
gnb\_id\_dummy & Integer & -- & Auxiliary dataset-internal gNB identifier (NR only, where available) \\
sector\_id & Integer & -- & Sector identifier within the site \\
timestamp / time & datetime & -- & Measurement time in UTC (\texttt{datetime64[ns]}) \\
latitude & Float & deg & Geographic latitude (EPSG:4326) \\
longitude & Float & deg & Geographic longitude (EPSG:4326) \\
\hline
\end{tabular}
\end{table}

\subsection*{Radio-Layer Metrics}

In addition to these identifiers, all measurement datasets (scanner and phone)
include physical-layer signal and quality indicators. These metrics are defined
consistently across LTE and NR where applicable and are summarized in
\autoref{tab:radio_metrics}.

\begin{table}[!t]
\caption{Radio-layer signal and quality indicators.}
\label{tab:radio_metrics}
\renewcommand{\arraystretch}{1.2}
\centering
\setlength{\tabcolsep}{3pt}
\begin{tabular}{|L{90pt}|L{40pt}|L{35pt}|L{275pt}|}
\hline
\textbf{Column} & \textbf{Type} & \textbf{Unit} & \textbf{Description} \\
\hline
rsrp\_dbm & Float & dBm & Reference Signal Received Power (RSRP) \\
rsrq\_db & Float & dB & Reference Signal Received Quality (RSRQ) \\
rssi\_dbm & Float & dBm & Received Signal Strength Indicator (RSSI) \\
sinr\_db & Float & dB & Signal-to-Interference-plus-Noise Ratio (SINR); reported for phone measurements and NR scanner measurements \\
cinr\_db & Float & dB & Carrier-to-Interference-plus-Noise Ratio (CINR); reported for LTE scanner measurements only \\
pathloss\_db & Float & dB & Path loss estimated from transmit reference power \\
\hline
\end{tabular}
\end{table}

\noindent\textbf{Note:} CINR is reported only for LTE scanner measurements. All other
signal-quality ratio values are stored as SINR (\texttt{sinr\_db}).

Finally, \autoref{tab:source_specific} summarizes additional fields that are
specific to either the measurement source (scanner vs.~phone) or the RAT (LTE vs.~NR). These include PDSCH/PUSCH throughput, timing advance,
and NR beam-level information.

\begin{table}[!t]
\caption{Source-specific and RAT-specific fields.}
\label{tab:source_specific}
\renewcommand{\arraystretch}{1.2}
\centering
\setlength{\tabcolsep}{3pt}
\begin{tabularx}{\columnwidth}{|l|l|l|l|X|}
\hline
\textbf{Column} & \textbf{Applies to} & \textbf{Type} & \textbf{Unit} & \textbf{Description} \\
\hline
beam\_index, beam\_type & 5G NR only & Integer, String & -- & Beam index and type (Serving or Detected) \\
dl\_throughput\_mbps & Phone only & Float & Mbit/s & Downlink PDSCH throughput \\
ul\_throughput\_mbps & Phone only & Float & Mbit/s & Uplink PUSCH throughput \\
timing\_advance & Phone only & Integer & -- & Timing advance (round-trip delay indicator) \\
\hline
\end{tabularx}
\end{table}

\subsection*{Estimated Cell Information}

This subdirectory provides metadata for LTE and 5G NR cells estimated from
decoded broadcast information, public deployment registers, and
measurement-based geolocation. Each record represents one estimated cell or
sector and includes stable identifiers, carrier frequency information,
geographic location, and sector configuration parameters such as antenna
azimuth and height.

Where standardized network-reported identifiers are unavailable (\textit{i.e.}, for NR
measurements collected in NSA mode), auxiliary dataset-internal
identifiers are used to enable consistent grouping of measurements and linkage
to the corresponding estimated cell metadata. Azimuth angles follow the
geographic convention, with $0^\circ$ corresponding to north and positive values
counted clockwise. The available columns are summarized in
\autoref{tab:cellinfo_cols}.

\begin{table}[!t]
\caption{Columns of the estimated cell information files.}
\label{tab:cellinfo_cols}
\renewcommand{\arraystretch}{1.2}
\centering
\setlength{\tabcolsep}{3pt}
\begin{tabular}{|l|l|l|l|}
\hline
\textbf{Column} & \textbf{Type} & \textbf{Unit} & \textbf{Description} \\
\hline
technology & String & -- & Radio access technology (LTE or NR) \\
enb\_id & Integer & -- & eNodeB identifier of the site (LTE) \\
sector\_id & Integer & -- & Sector identifier within the site \\
cell\_id & Integer & -- & Unique cell identifier (LTE) \\
cell\_id\_dummy & Integer & -- & Auxiliary dataset-internal NR cell identifier \\
gnb\_id\_dummy & Integer & -- & Auxiliary dataset-internal gNB identifier \\
pci & Integer & -- & Physical Cell Identity \\
channel\_number & Integer & -- & Radio frequency channel number \\
latitude & Float & deg & BS latitude (EPSG:4326) \\
longitude & Float & deg & BS longitude (EPSG:4326) \\
height\_m & Integer & m & Antenna height above ground level \\
azimuth\_deg & Integer & deg & Antenna azimuth orientation \\
\hline
\end{tabular}
\end{table}

The use of consistent column naming and units facilitates merging cell
information with corresponding scanner and phone measurements, enabling
spatially and temporally aligned analyses of both coverage and performance.

\subsection*{Vienna City Model}

This subdirectory provides environmental data forming the basis of the Vienna
DNT. It includes the Building Model (BKM) and Ground-Level Model
(GLM), both stored as GeoTIFF rasters in EPSG:31287 and compressed using 7-Zip.
Both rasters provide height information at \SI{1}{\metre} spatial resolution,
representing building heights and terrain elevation, respectively. Together,
these layers enable environment-aware analyses and learning tasks such as
(i) training propagation surrogates conditioned on local urban morphology,
(ii) learning blockage- and diffraction-aware coverage models, or
(iii) supporting AI-assisted network planning pipelines that combine cell
metadata with 3D urban context.

\section*{Technical Validation}

This section summarizes the validation analyses performed to assess the
reliability and accuracy of the Vienna~4G/5G Dataset. The focus lies on aspects
that are essential for downstream use in research, simulation, and data-driven
modeling, namely the spatial accuracy of the recorded measurement locations,
the consistency of cell assignments in the scanner data, and the statistical
properties of KPIs such as RSRP and RSRQ.
In addition, the specifications and accuracy limits of the employed measurement
equipment are summarized to provide context for the achievable precision of
the recorded data. Where appropriate, residual distributions, box plots, and
summary tables are used to illustrate errors and variances observed in the
dataset.

\subsection*{Georeferencing Accuracy}

All measurement locations in the released dataset are stored as raw GPS records
in geographic coordinates (EPSG:4326, latitude and longitude). For
distance-based analyses, such as spatial error evaluation or map-based
snapping, reprojection to a metric coordinate reference system is required; in
this work, EPSG:31287 was used. The reprojected coordinates are not included in
the dataset itself, but all positioning accuracy results reported in this
section were obtained after reprojection.

GPS measurements are subject to noise, particularly in dense urban
environments with limited satellite visibility, and recorded positions may
deviate from the actual road geometry. To quantify this effect, GPS samples
were compared to the nearest street segments extracted from OpenStreetMap. Let
$p_i \in \mathbb{R}^2$ denote a reprojected GPS point and let $\mathcal{S}$ be
the set of street polylines. The snapped point and corresponding lateral offset
are defined as
\begin{equation*}
\begin{aligned}
s(p_i) &= \operatorname*{arg\,min}_{q \in \mathcal{S}} \lVert p_i - q \rVert_2,\\
d_i    &= \lVert p_i - s(p_i) \rVert_2 .
\end{aligned}
\end{equation*}
The overall positioning error was evaluated using the root mean square error
(RMSE),
\begin{equation*}
\mathrm{RMSE} = \sqrt{\frac{1}{N} \sum_{i=1}^{N} d_i^2},
\end{equation*}
where $d_i$ denotes the per-sample horizontal offset and $N$ is the total
number of samples.

\begin{figure}
\centerline{\includegraphics[width=0.7\textwidth]{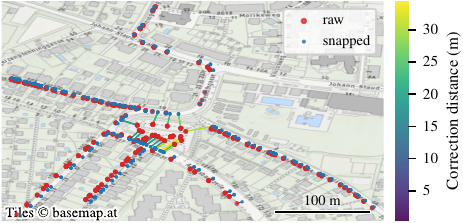}}
\caption{Example of raw GPS samples (red) and corresponding snapped positions (blue) on the nearest road, with correction vectors shown as colored lines. The line color encodes the correction distance in meters.\label{fig:gps_snap_segment}}
\end{figure}

Applied to the Vienna~4G/5G Dataset, this analysis yields an RMSE of
approximately $\SI{4}{\metre}$. This value is consistent with expected GPS
accuracy under urban driving conditions and confirms that the recorded
locations provide sufficient spatial precision for drive-test--based
propagation analysis and BS localization. A representative example
illustrating raw GPS samples, snapped positions, and the corresponding
correction vectors is shown in \autoref{fig:gps_snap_segment}.

\subsection*{Cell Assignment Consistency}

The consistency of assigning physical-layer scanner measurements to broadcast
system information was evaluated by analyzing temporal and spatial residuals
between CRS and SSB records and their MIB and
SIB decoding events. This analysis provides an internal
validation of the scanner parsing and enrichment pipeline.

Candidate assignments were identified based on exact matching keys
(PCI, channel number, measurement file) and subsequently evaluated using
temporal and spatial consistency criteria. The nominal acceptance set was
defined as
\begin{equation*}
\begin{aligned}
\mathcal{A}
  &= \bigl\{(i,j):\; |\Delta t_{ij}| \le \tau_t \ \wedge\ \Delta x_{ij} \le \tau_x \bigr\},\\
\tau_t &= \SI{10}{\second}, \qquad \tau_x = \SI{750}{\metre},
\end{aligned}
\end{equation*}
where $\Delta t_{ij}$ denotes the timestamp difference between a CRS/SSB detection and a MIB/SIB decoding event, and $\Delta x_{ij}$ the corresponding point-to-point
distance evaluated in EPSG:31287.

Each CRS/SSB record is linked to at most one candidate MIB/SIB event. Records with at least one candidate (\textit{i.e.}, at least one event sharing the exact matching keys PCI, channel number, and measurement file) are classified as ok if both criteria are satisfied, or as time-outlier and/or space-outlier if the temporal and/or spatial criterion is violated. Records with no candidate under the exact-key match are labeled unmatched. For compactness, \autoref{tab:matching_stats_a} reports matched vs. unmatched counts, where matched includes ok and any outlier case (including violations of both criteria). \autoref{tab:matching_stats_b} reports the percentages of time-outlier, space-outlier, and unmatched samples; the fraction of ok samples equals 100\,\% minus the reported percentages.

Across the full dataset, 2\,649\,377 CRS/SSB records were evaluated, of which
472\,645 (\SI{17.84}{\percent}) could not be matched to a corresponding MIB/SIB decoding event. In LTE,
655\,310 records were processed, with \SI{14.90}{\percent} unmatched, whereas in
5G NR, 1\,994\,067 records were evaluated, with \SI{18.81}{\percent} unmatched.

\begin{table}[t]
\caption{CRS/SSB to MIB/SIB assignment outcomes split by technology: counts of matched and unmatched records. Matched includes ok, time-outlier, and space-outlier.}
\label{tab:matching_stats_a}
\renewcommand{\arraystretch}{1.2}
\centering
\setlength{\tabcolsep}{5pt} 

\begin{tabular}{|l|l|l|l|}
\hline
Technology & $N$ total & Matched (any) & Unmatched \\
\hline
Overall & 2\,649\,377 & 2\,176\,732 & 472\,645 \\
LTE     &   655\,310 &   557\,668 &  97\,642 \\
5G NR   & 1\,994\,067 & 1\,619\,064 & 375\,003 \\
\hline
\end{tabular}
\end{table}

\begin{table}[t]
\caption{CRS/SSB to MIB/SIB assignment outcomes split by technology: percentages of time-outlier, space-outlier, and unmatched records. The fraction of ok samples equals $100\,\%$ minus the reported percentages.}
\label{tab:matching_stats_b}
\renewcommand{\arraystretch}{1.2}
\centering
\setlength{\tabcolsep}{5pt}

\begin{tabular}{|l|l|l|l|}
\hline
Technology & Time-outlier [\%] & Space-outlier [\%] & Unmatched [\%] \\
\hline
Overall & 13.84 & 0.53 & 17.84 \\
LTE     & 17.99 & 1.96 & 14.90 \\
5G NR   & 12.48 & 0.06 & 18.81 \\
\hline
\end{tabular}
\end{table}

\begin{figure}[!t]
\centering
\includegraphics[width=5.0in]{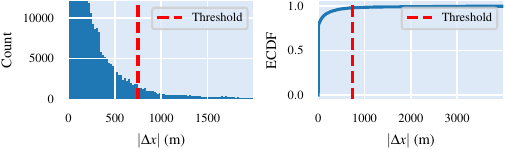}
\caption{Spatial residuals for CRS/SSB to MIB/SIB assignment. Histogram (clipped
at 12\,000) and empirical CDF of \boldmath{$|\Delta x|$}. The dashed vertical line
denotes the acceptance threshold
\boldmath{$\tau_x=\SI{750}{\metre}$}.}
\label{fig:delta_x_plot}
\end{figure}

Temporal residuals exhibit a pronounced long tail,
\begin{equation*}
\begin{aligned}
\operatorname{median}\,|\Delta t| &= \SI{0}{\second},\\
P_{90} &= \SI{2030}{\second},\quad
P_{95} = \SI{4585}{\second},\quad
P_{99} = \SI{14763}{\second},
\end{aligned}
\end{equation*}
reflecting asynchronous decoding behavior and buffering effects in the scanner
logs. In contrast, spatial residuals are considerably tighter
(\autoref{fig:delta_x_plot}),
\begin{equation*}
\begin{aligned}
\operatorname{median}\,|\Delta x| &= \SI{0}{\metre},\\
P_{90} &= \SI{145}{\metre},\quad
P_{95} = \SI{342}{\metre},\quad
P_{99} = \SI{1\,423}{\metre}.
\end{aligned}
\end{equation*}

\begin{figure}[!t]
\centerline{\includegraphics[width=5.0in]{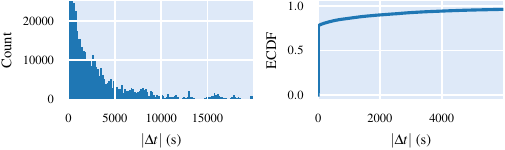}}
\caption{Temporal residuals for CRS/SSB to MIB/SIB assignment. Histogram (clipped
at 25\,000) of \boldmath{$|\Delta t|$} and ECDF of
\boldmath{$|\Delta t|$}.}
\label{fig:delta_t_plot}
\end{figure}

Because the nominal temporal threshold $\tau_t=\SI{10}{\second}$ proved too
restrictive given the observed jitter, the released dataset relies on a spatial
fallback strategy. For each CRS/SSB record $i$, the MIB/SIB record $j^\ast$ that
minimizes the spatial residual is selected,
\[
j^\ast(i) = \arg\min_{j \in \mathcal{C}(i)} \Delta x_{ij}
\quad \text{s.t.} \quad \Delta x_{ij} \le \tau_x,
\]
where $\mathcal{C}(i)$ denotes the set of candidate MIB/SIB decoding events consistent
with the CRS/SSB key tuple. Temporal residual statistics are nevertheless
reported for transparency and may inform alternative threshold choices in
downstream analyses (\autoref{fig:delta_t_plot}).

\subsection*{Key Performance Indicator Statistics}

\begin{figure}[!t]
\centerline{\includegraphics[width=3.5in]{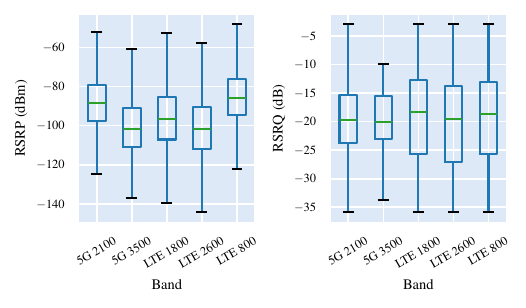}}
\caption{Distributions of RSRP and RSRQ across LTE and NR carriers. Boxplots are
grouped by carrier frequency: LTE~800~MHz (B20), LTE~1800~MHz (B3),
LTE~2600~MHz (B7), NR~2100~MHz (n1), and NR~3500~MHz (n78).}
\label{fig:kpi_boxplots_freq}
\end{figure}

The distributions of RSRP and RSRQ derived from the scanner-based measurements
across the available LTE and NR carriers are summarized in
\autoref{fig:kpi_boxplots_freq}. As expected, the LTE \SI{800}{\mega\hertz}
low-band carrier (B20) exhibits the strongest received power, reflecting
favorable propagation conditions at lower frequencies. Mid-band LTE
\SI{1800}{\mega\hertz} (B3) and \SI{2600}{\mega\hertz} (B7) show progressively
lower RSRP levels, while the NR \SI{3500}{\mega\hertz} C-band (n78) displays the
weakest received power due to increased path loss and reduced building
penetration. The NR \SI{2100}{\mega\hertz} carrier (n1) exhibits intermediate
behavior, consistent with its frequency range.

In terms of RSRQ, low-band LTE again provides the most stable quality values,
with median levels close to $\SI{-9}{\decibel}$. Mid- and high-band layers show a
wider dispersion of RSRQ, attributable to higher cell density and increased
interference. Overall, these band-dependent differences confirm that the
scanner measurements capture realistic propagation and interference
characteristics across heterogeneous spectrum deployments.

\begin{figure}[!t]
\centerline{\includegraphics[width=\columnwidth]{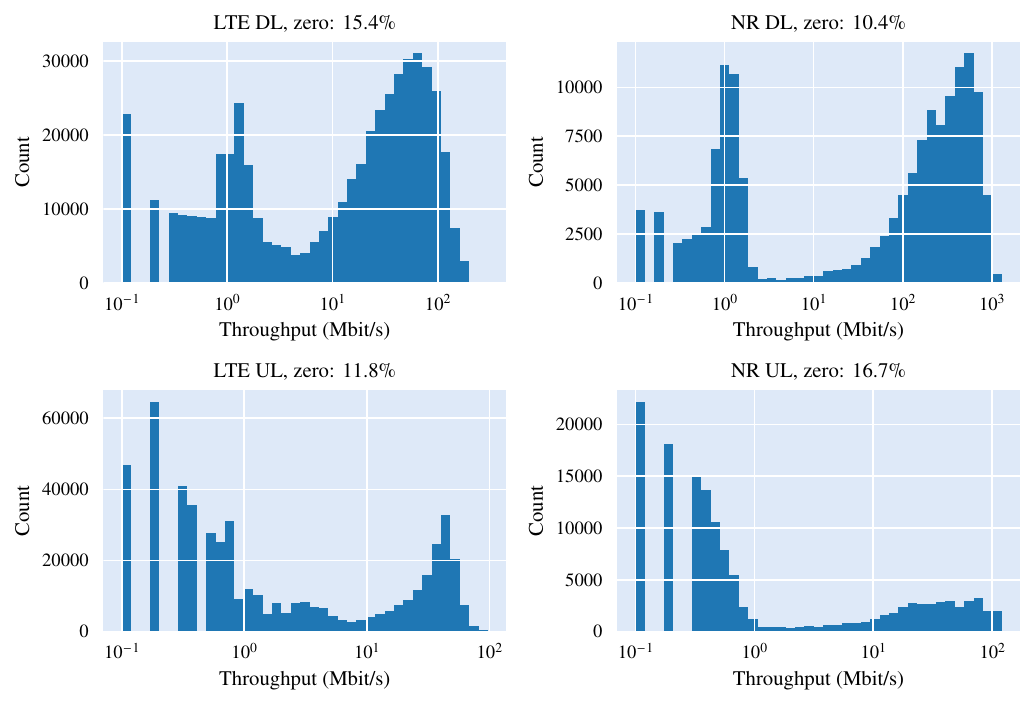}}
\caption{Distribution of non-zero PDSCH and PUSCH throughput samples in the released phone LTE (left) and NR (right) datasets. Panel titles indicate the fraction of zero-throughput samples.}
\label{fig:throughput_distribution}
\end{figure}

\autoref{fig:throughput_distribution} shows the distributions of non-zero downlink and
uplink PDSCH/PUSCH throughput values in the released phone LTE and NR datasets.
A substantial fraction of zero or near-zero PDSCH/PUSCH throughput samples is
expected in this setup. The phone tables represent continuous diagnostic logs
collected during mobile speed-test operation, and the UE is not necessarily
scheduled on the downlink or uplink shared channel in every sampling interval.
Near-zero values may therefore occur during intervals between active transfer
phases, connection interruptions, handovers, radio-link recovery, poor coverage,
uplink grant limitations, or other lower-layer scheduling states. These samples
were retained because they reflect the actual instantaneous lower-layer
transmission activity during the drive tests. For throughput-focused analyses,
users may filter active-transfer samples or analyze zero and non-zero samples
separately, depending on the intended application.

\subsection*{Plausibility and Validation of Estimated Base Station Metadata}

To support the creation of a DNT, a subset of BSs
was localized using timing-advance--based methods. For LTE, where network-reported cell identifiers enable unambiguous grouping and reference metadata
are available for part of the deployment, the estimated BS locations
and sector orientations were quantitatively assessed.

Localization accuracy was evaluated using the RMSE
between estimated and reference BS positions, while sector
orientations were assessed using the mean absolute circular error between
estimated and reference azimuth angles. Across the validated LTE sites, a
localization RMSE of $\SI{38.55}{\metre}$ and a mean absolute circular error of
$\SI{27.95}{\degree}$ were observed, indicating a physically consistent and
plausible reconstruction of the underlying deployment.

For 5G NR, handset-based measurements were collected in NSA
mode, where physical NR cell or gNB identifiers are not exposed by commercial
devices. As a result, measurements cannot be grouped unambiguously by base
station, and external reference metadata are not available for direct
quantitative validation. Consequently, the assessment of estimated 5G BS metadata is limited to internal consistency and plausibility
checks.

\begin{figure}[!t]
\centering
\includegraphics[width=0.7\textwidth]{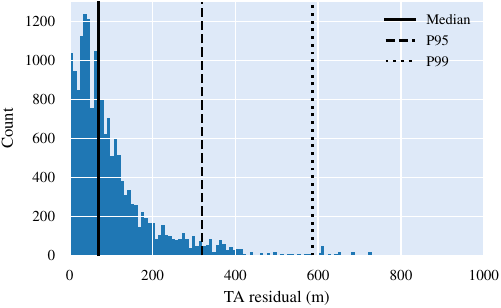}
\caption{Timing-advance residual distribution for 5G NR. Vertical lines indicate
the median, 95th percentile, and 99th percentile of the residuals. For clarity,
the horizontal axis is limited to 1000\,m.}
\label{fig:nr_ta_residual}
\end{figure}

TA consistency was evaluated by comparing the TA-derived range
to the geometric distance between each UE measurement position and the
corresponding estimated BS location. With an NR TA step size of
$1~\text{TA}=\SI{1.2216}{\metre}$, the per-sample residual is defined as
\begin{equation*}
r_i=\bigl|\lVert \mathbf{x}_i-\hat{\mathbf{b}}\rVert_2 - 1.2216\cdot
\text{TA}_i\bigr|,
\end{equation*}
where $\mathbf{x}_i$ denotes the UE position and
$\hat{\mathbf{b}}$ the estimated BS position. As shown in
\autoref{fig:nr_ta_residual}, the resulting residual distribution exhibits a
median of $\SI{69.6}{\metre}$ with $P_{95}=\SI{318.8}{\metre}$ and
$P_{99}=\SI{585.4}{\metre}$ across $N=17{\,}433$ NR samples, indicating that the
inferred 5G locations remain compatible with the underlying TA
constraints.

\subsection*{Scanner Specifications}

Passive measurements were collected using a PCTEL~IBflex scanning receiver
connected to two OmniLOG~PRO omnidirectional antennas and a dedicated GPS
receiver. \autoref{tab:scanner_specs} summarizes the key specifications that are
relevant for interpreting the dataset accuracy. Further details are available
in the manufacturer's datasheet~\cite{ibflex_datasheet}. The commercial
handsets used for phone-based measurements are treated as representative user
equipment rather than calibrated measurement instruments; detailed device
information is provided in the dataset documentation.

\begin{table}[!t]
\caption{Scanner measurement specifications for LTE and 5G NR modes.}
\label{tab:scanner_specs}
\renewcommand{\arraystretch}{1.2}
\centering
\setlength{\tabcolsep}{3pt}

\begin{tabular}{|l|l|l|}
\hline
\multirow{2}{*}{} & \multicolumn{2}{c|}{\textbf{Specification}$^{a}$} \\
\hhline{|~|-|-|}
 & \multicolumn{1}{c|}{\textbf{LTE}} & \multicolumn{1}{c|}{\textbf{NR}} \\
\hline
Measurement mode & \multicolumn{2}{L{130pt}|}{Synchronization channels} \\
\hline
Reported data & \multicolumn{2}{L{130pt}|}{PCI, RSRP, RSRQ, CINR (LTE), SINR (NR), delay spread, time, latitude, longitude; NR additionally: beam index} \\
\hline
Max.\ number of PCIs & 16 & 8 (max.\ 8 beams/PCI) \\
\hline
Subcarrier spacing & 15~kHz & 15/30~kHz \\
\hline
Minimum detection level & $-140$~dBm & $-135$~dBm \\
\hline
Accuracy (CINR/SINR) & $\pm 1$~dB & $\pm 2$~dB \\
\hline
Device measurement rate (typical) & \SI{50}{\per\second} & \SI{30}{\per\second} \\
\hline
\end{tabular}

\vspace{4pt}
\parbox{\linewidth}{\footnotesize
$^{a}$The reported values represent nominal device specifications. In particular, the stated measurement rates do not correspond to the temporal granularity of the released dataset; the recorded reference-signal measurements have an effective temporal granularity of approximately 500\,ms (see Methods).
}
\end{table}


\section*{Usage Notes} 
All timestamps in the dataset are stored as \texttt{datetime64[ns]} values
without explicit timezone information and are expressed in Coordinated
Universal Time (UTC). Users working with local time representations should
apply the appropriate timezone conversion if required for visualization or
interpretation.

Measurement files containing spatial information include raw GPS coordinates in
EPSG:4326. This geographic coordinate reference system is used for storage because
it corresponds to the native latitude/longitude representation of GPS measurements
and is widely interoperable with common mapping and geospatial software. However,
EPSG:4326 represents positions in angular units rather than in meters, so
Euclidean distances, buffers, nearest-neighbor searches, and raster-window
operations should not be performed directly in this coordinate system. For
distance-based analyses, spatial filtering, and alignment with the Vienna city
model, we therefore recommend transforming the data to EPSG:31287, a projected
metric coordinate reference system suitable for Austria. The building and terrain
rasters provided with the dataset are already stored in EPSG:31287, which avoids
unnecessary reprojection when combining the measurement data with the city-model
layers. To improve spatial consistency in map-based analyses, users may optionally
snap measurement locations to the nearest street segments; however, the dataset
itself does not apply any map-matching or spatial correction.

All handset 5G~NR measurements (phone data) in this dataset were collected in NSA
mode, with LTE serving as the primary connection. As a result, physical NR cell
identifiers (\textit{i.e.}, \texttt{cell\_id}, \texttt{gnb\_id}, \texttt{sector\_id}) are
not reported by the handset logging interface and are therefore not present in
the NR phone traces. Only LTE anchor cell identifiers are available in the phone
dataset. NR and LTE samples can be temporally aligned to associate each NR
measurement with its corresponding LTE anchor, enabling joint LTE--NR analyses
at the connection level.

Where available, auxiliary dataset-internal identifiers
(\texttt{cell\_id\_dummy}, \texttt{gnb\_id\_dummy}) can be used to associate a subset
of NR measurements with inferred cell characteristics and estimated cell metadata.
These identifiers are not standardized network identifiers and may be null.

Weather parameters are not included directly in the dataset. Users interested in
weather-dependent propagation effects can retrieve corresponding historical
weather information from public meteorological data services such as the
Open-Meteo Historical Weather API
(\url{https://open-meteo.com/en/docs/historical-weather-api}), using the
measurement timestamps and coordinates provided in this dataset. The API provides
hourly weather variables, including temperature, relative humidity, precipitation,
pressure, cloud cover, wind speed, and wind direction. An example of retrieving
and aligning such weather data is provided in the accompanying demo notebook.

Finally, users should be aware that the dataset reflects real-world drive-test
conditions. Measurements may therefore include transient effects such as
handover dynamics, beam switching, and short-term fluctuations in radio
conditions. These characteristics make the dataset particularly suitable for
data-driven and learning-based analyses but may require appropriate filtering
or aggregation depending on the intended application.

\section*{Data Availability}
The dataset is archived on Zenodo under the concept DOI 
\url{https://doi.org/10.5281/zenodo.18322065}. 
This article describes version 5, which is available under the versioned DOI 
\url{https://doi.org/10.5281/zenodo.21372657}~\cite{wiedner_vienna_2025}.

\section*{Code Availability} 

Custom code generated for this study is available at
\url{https://squid.nt.tuwien.ac.at/gitlab/vienna-4g-5g-dataset/vienna-4g5g-drive-test-dataset-code}. The repository contains public-release versions of the LTE and NR scanner and handset processing
pipelines used for the Vienna~4G/5G Drive-Test Dataset, including timestamp
conversion, event extraction, record harmonization, operator anonymization,
cell/beam-level processing, and generation of publication-ready tables. The
repository also includes a demo notebook illustrating common operations on the
released dataset, including loading and merging dataset tables, transforming
coordinates to EPSG:31287, extracting and loading the Vienna building and
terrain rasters, and retrieving and merging corresponding historical weather
data from Open-Meteo.

The raw proprietary Nemo Outdoor export files and campaign-specific export
configurations are not included in the repository or dataset. Consequently, the
raw-export processing scripts document the dataset-generation workflow but are
not directly executable from the released dataset alone. 

\section*{Acknowledgements}
The authors gratefully acknowledge Stadtvermessung Wien for providing building and terrain information of the city of Vienna, which forms an integral part of the DT included in this dataset. 

\section*{Funding}
This work was supported through the Christian Doppler Laboratory for Digital Twin assisted AI for Sustainable Radio Access Networks (Institute of Telecommunications, TU Wien). Funding for the laboratory is provided by the Austrian Federal Ministry of Economy, Energy and Tourism, the National Foundation for Research, Technology and Development, and the Christian Doppler Research Association. The authors acknowledge the TU Wien Bibliothek for financial support through its Open Access Funding Programme.

\printbibliography

\end{document}